\documentstyle[epsfig]{aipproc}

\begin{document}
\title{The Equilibrium Structure of Cosmological 
Halos}

\author{Ilian T. Iliev$^*$ and Paul R. Shapiro$^{\dagger}$}
\address{$^*$IA-UNAM, Mexico\\
$^{\dagger}$The University of Texas at Austin}

\maketitle

\begin{abstract}
We have derived an analytical 
model for the postcollapse equilibrium structure of 
cosmological halos as nonsingular truncated isothermal spheres (TIS)
and compared this model with observations and simulations of cosmological
halos on all scales. Our model is in good agreement with the
observations of the internal structure of dark-matter-dominated halos from 
dwarf galaxies to X-ray clusters. It 
reproduces many of the average properties of halos in CDM simulations to
good accuracy, including the density profiles outside the central region, while
avoiding the possible discrepancy at small radii between observed 
galaxy and cluster density profiles and the singular density profiles 
predicted by 
N-body simulations of the CDM model. 
While much attention has been focused lately on this possible discrepancy, 
we show that the observed galaxy rotation curves and correlations of 
halo properties
nevertheless contain valuable additional information with which to
test the theory, despite this uncertainty at small radii. The available data 
allows us to constrain the fundamental cosmological parameters and also to put 
a unique constraint on the primordial density fluctuation power spectrum at
large wavenumbers (i.e. small mass scale).
\end{abstract}

{\bf The TIS Model.} Our model is described in detail in \cite{SIR} for an EdS
universe and generalized to a low-density universe, either matter-dominated or
flat with $\Lambda>0$ in \cite{ISb}.
An initial top-hat density perturbation collapses and virializes, 
which leads to a nonsingular TIS in 
hydrostatic equilibrium, a solution of the Lane-Emden 
equation (appropriately modified for $\Lambda\neq 0$). 
Using the anzatz that the resulting TIS sphere is the one with the
minimum-energy, out of the family of possible solutions, we find that a 
top-hat perturbation 
collapse leads to a unique, nonsingular TIS, yielding a universal, 
self-similar density profile for the  postcollapse equilibrium of cosmic 
halos. Our solution  has a unique length scale and amplitude set by the top-hat 
mass and collapse epoch, with a density proportional to the background density
at that epoch. The density profiles for gas and dark matter are
assumed to be the same. 

{\bf Rotation Curves of Dark-Matter Dominated Galactic Halos.}
The TIS profile matches the observed mass profiles of 
dark-matter-dominated dwarf galaxies, which are well-fit by the empirical
density profile of \cite{B}, with a finite density core. 
The TIS profile gives a nearly perfect fit to the Burkert profile,
providing it with theoretical underpinning and 
a cosmological context \cite{ISa}.
\begin{figure}
\epsfig{file=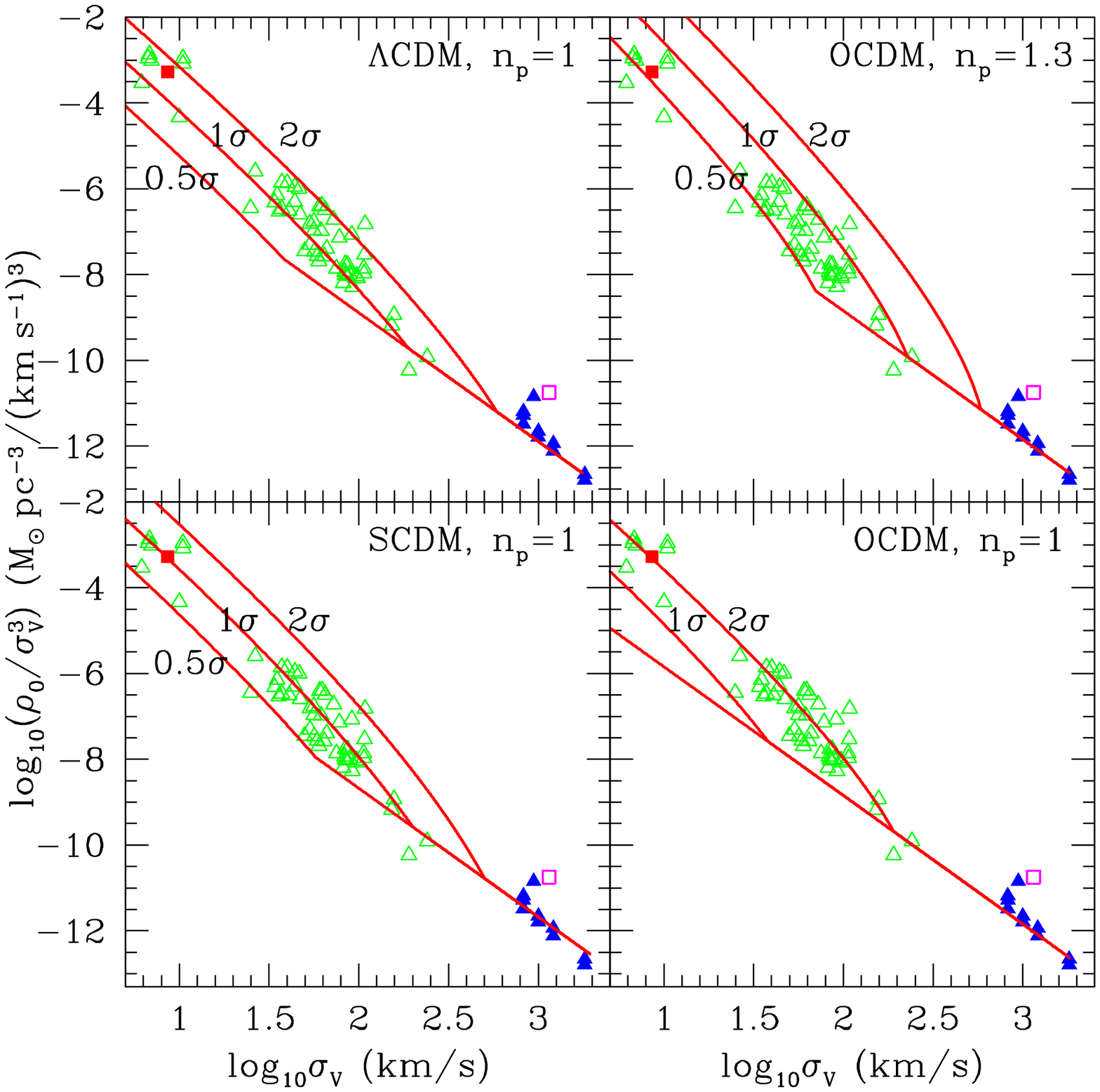,height=2.7in}
\epsfig{file=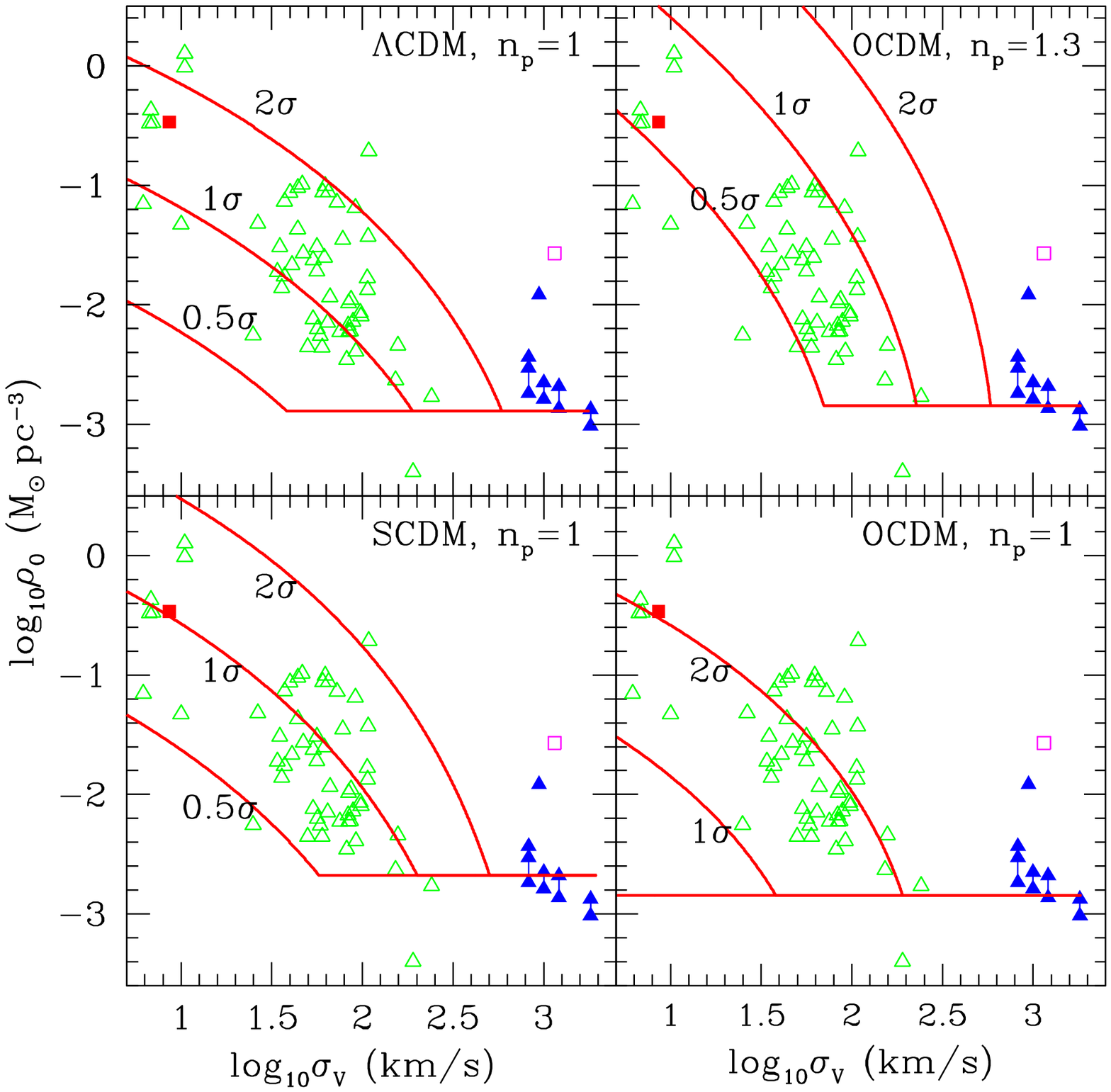,height=2.7in}
\vspace{10pt}
\caption{
Galaxy and cluster halo phase-space density 
$Q\equiv\rho_0/\sigma_V^3$ (left panels) and halo central density 
$\rho_0$ (right panels) versus the halo velocity dispersion $\sigma_V$:
empty triangles = galaxy data from [9]; filled square =
data for dwarf galaxy Leo I from [10];
filled triangles = cluster data from [2] (the 
filled triangles with the same $\sigma_V$ correspond to different 
mass estimates 
for the same cluster); empty square = galaxy cluster 
CL 0024 ($\sigma_V$ is from [3], central density 
obtained in [13]
by fitting TIS profile to the strong lensing data of [15]);
curves = TIS + Press-Schechter (PS) prediction for four popular cosmological 
models, as labelled ($n_p$ is the tilt of the power spectrum). 
The models in the two upper panels are COBE- and 
cluster-normalized, with $\Omega_0=0.3$ and $\lambda_0=0.7$ and 0, 
respectively. Results are for fluctuations of 
different amplitudes $\nu \equiv {\delta_{\rm crit}}/{\sigma(M)}$, 
where $\delta_{\rm crit}$ is the value of the linear density 
contrast with respect to the background density for a tophat fluctuation,
extrapolated to the time when the actual nonlinear density inside the tophat
reaches infinity, as labelled with $\nu-\sigma$. Curves for
each $\nu$ connect to the curve for $z_{\rm coll}=0$,
for those $\nu-\sigma$ fluctuations which have not yet collapsed by 
$z=0$.
}
\label{fig1}
\end{figure}

We have also combined the TIS halo model with the Press-Schechter (PS)
formalism, which 
predicts the typical collapse epoch for objects of a given mass 
in the CDM model, to explain the observed 
correlation of $v_{\rm max}$ and 
$r_{\rm max}$ for dwarf spiral and LSB galaxies.
The observational data indicates preference for the
currently-favored, flat $\Lambda$CDM universe 
($\Omega_0=1-\lambda_0=0.3$, $h=0.7$).
For more details on our methods and results, see \cite{ISa}.

{\bf The Correlations of Halo Core and 
Maximum Phase-Space Densities with Velocity
Dispersion}. A comparison of observed halo properties with other
correlations predicted by this TIS+PS approach can test the CDM model
while constraining the fundamental cosmological parameters and the 
power-spectrum shape at small mass scales (i.e. large wavenumbers $k$).
The core densities $\rho_0$ and maximum phase-space densities 
$Q\equiv\rho_0/\sigma_V^3$ for dark-matter dominated halos are
predicted to be correlated with their
velocity dispersions $\sigma_V$ as shown in
Figure~1. For cold, collisionless DM, $Q$ is expected to be
almost independent of the effects of baryonic dissipation \cite{S}.
The data on halos from dwarf spheroidal to clusters is consistent with
these predictions, with preference for the flat, $\Lambda$CDM model.
There have been recent claims that $\rho_0=$const for all cosmological halos,
independent of their mass, and that such behavior is expected for 
certain types of SIDM
\cite{FDCHA,KKT}.
This claim, however, does not seem to be supported by the current data 
(Figure 1). 

{\bf Galaxy Clusters: TIS vs. CDM Simulations}
We have shown previously \cite{IS,SI,SIR}
that the TIS halo model predicts to great accuracy the 
internal structure of X-ray clusters found by gas-dynamical/N-body 
simulations of cluster formation in the CDM model at $z=0$. 
The TIS prediction for the redshift evolution
of the halo mass-temperature and mass-velocity 
dispersion relations for galaxy clusters
also matches to high accuracy ($\sim$ few percent) the empirical relations
derived in \cite{ME} from CDM gas/N-body
simulations and by the Virgo Consortium from
their Hubble volume N-body simulations [Evrard, private communication].

\centerline{\bf{Acknowledgments}}

This research was supported by NSF grant INT-0003682 from 
the International Research Fellowship Program and the Office of 
Multidisciplinary Activities of the Directorate for Mathematical and
Physical Sciences to ITI
and grants NASA ATP NAG5-7363 and NAG5-7821, NSF ASC-9504046, and Texas 
Advanced Research Program 3658-0624-1999 to PRS.


\begin{references}
\bibitem{B} Burkert, A. ApJ, {\bf 447}, L25 (1995).
\bibitem{DH} Dalcanton, J.J., and Hogan, C.J., ApJ, submitted (astro-ph/0004381)
\bibitem{DSPBCEO} Dressler, A., Smail, I., Poggianti, B.M., Butcher, H., Couch, W.J.,
		Ellis, R.S., and Oemler, A., Jr. ApJS, {\bf 122}, 51 (1999).
\bibitem{FDCHA} Firmani, C., D'Onghia, E., Chincarini, G., 
		Hernandes, X., and Avila-Reese, V. MNRAS, 321, 713 (2000)
\bibitem{IS} Iliev I.T., and Shapiro P.R., in "The Seventh Texas-Mexico 
	Conference on Astrophysics: Flows, Blows, and Glows," eds. W. Lee and S.
	Torres-Peimbert, RevMexAA (Serie de Conferencias), in press (2001) (astro-ph/0006184) 
\bibitem{ISa} Iliev, I.T., and Shapiro, P.R., ApJ, {\bf 546}, L5 (2001a).
\bibitem{ISb} Iliev, I.T., and Shapiro, P.R., MNRAS, in press (2001b)
(astro-ph/0101067).
\bibitem{KKT} Kaplinghat, M., Knox, L., and Turner, M.S., preprint (astro-ph/0005210) 
\bibitem{KF} Kormendy, J. \& Freeman K.C. 2001, in preparation.
\bibitem{MOVK} 
	Mateo, M., Olszewski, E.W., Vogt, S.S., and Keane, M.J. ApJ, {\bf 116}, 2315 (1998)
\bibitem{ME} Mathiesen, B.F., Evrard, A.E. ApJ, 546, 100 (2001)
\bibitem{S} Sellwood, J.A. ApJ, 540, 1L (2000)
\bibitem{SI} Shapiro P.R., and Iliev I.T. ApJ, {\bf 542}, L1 (2000)
\bibitem{SIR}Shapiro, P.R., Iliev, I.T., and Raga, A.C., MNRAS {\bf 307}, 203 (1999).
\bibitem{TKD} Tyson, J.A., Kochanski, G.P., and Dell'Antonio, I.P. ApJ, 
	{\bf  498}, L107 (1998).
\end{references}
\end{document}